\RequirePackage[hyphens]{url}
\documentclass[sigconf,techreport]{acmart}

\pdfoutput=1

\makeatletter
\@ifclasswith{acmart}{techreport}{
  \settopmatter{printacmref=false}
  \renewcommand\footnotetextcopyrightpermission[1]{}
  \fancyhead{}
  \fancyfoot{}
  \fancyhead[L]{\shorttitle}
  \fancyhead[R]{\shortauthors}
  \fancyfoot[C]{\thepage}
}{}
\makeatother

\begin{document}

\title{Supporting Web Archiving via Web Packaging}
\author{Sawood Alam$^1$, Michele C Weigle$^1$, Michael L Nelson$^1$, Martin Klein$^2$, and Herbert {Van de Sompel}$^3$}
\affiliation{%
  \institution{$^1$Department of Computer Science, Old Dominion University, Norfolk, Virginia, USA}
}
\email{{salam,mweigle,mln}@cs.odu.edu}
\affiliation{%
  \institution{$^2$Los Alamos National Laboratory, New Mexico, USA}
}
\email{mklein@lanl.gov}
\affiliation{%
  \institution{$^3$Data Archiving and Networked Services, Netherlands}
}
\email{herbert.van.de.sompel@dans.knaw.nl}

\renewcommand{\shortauthors}{S. Alam et al.}
\renewcommand{\pdfmetaauthors}{Sawood Alam, Michele C. Weigle, Michael L. Nelson, Martin Klein, and Herbert {Van de Sompel}}

\begin{abstract}
We describe challenges related to web archiving, replaying archived web resources, and verifying their authenticity.
We show that \emph{Web Packaging} has significant potential to help address these challenges and identify areas in which changes are needed in order to fully realize that potential.
\end{abstract}

\maketitle

\section{Introduction}

\fancypagestyle{firstpagenote}{
  \fancyhf{}
  \fancyfoot[L]{\rule{0.66in}{0.4pt}\\ \footnotesize \emph{$^*$This is a position paper accepted at the ESCAPE Workshop 2019.} \url{https://www.iab.org/activities/workshops/escape-workshop/}}
}
\thispagestyle{firstpagenote}

Web archiving is the practice of preserving representations of web resources to enable replaying them in the future as accurately as they were at the time of capture.
A web archive can be seen as a datetime indexed caching proxy server that preserves every transaction (including responses that traditional proxy servers would be asked not to cache) indefinitely, allowing future replay.
Depending on the capabilities of available tools, computing and storage resources, curatorial workforce, intended use cases, and objectives, a web archive may choose to preserve complete \emph{HTTP} \emph{Request} and \emph{Response} messages, \emph{DNS} resolutions, \emph{TLS} exchanges, and other provenance metadata for each observation.
Two standards are widely used in web archiving: \emph{WARC files}~\cite{warcformat,warcintro} for preservation of observations and the \emph{Memento protocol}~\cite{memento:rfc} for datetime content negotiation at replay.
Most web archives preserve these resources in the well-established \emph{ISO} standard \emph{WARC} file format, which is a container file for an arbitrary number of records and metadata.
\emph{WARC} is somewhat like \emph{tarball}, but for web archiving, in which individual \emph{HTTP} transactions and other record types are prefixed with \emph{HTTP}-like \emph{WARC headers} for additional metadata and record length (for framing).

The \emph{Memento} protocol specifies time-based content negotiation so that a user-agent can request a representation of an original \emph{resource URI} (or \emph{URI-R}) at or close to a given time through intermediation of a \emph{TimeGate resource} (or \emph{URI-G}).
This past version can be from a version-aware origin server itself (e.g., \emph{Git} and \emph{Wiki}) or an observation recorded by a third party web archive (e.g., \url{archive.org}, \url{archive.is}, and \url{perma.cc}).
A \emph{memento} is a timestamped archived version of a resource representation that can be retrieved from a \emph{memento URI} (or \emph{URI-M}).
A \emph{composite memento} is an archived web page along with all its page requisites observed in a small temporal window close to the primary web page that are necessary for its proper rendering and meaningful interactions.

\emph{Web Packaging} is an emerging standard~\cite{webpackaging} that enables content aggregators and distributors to deliver related groups of resources from various origins to user-agents in the form of a package on behalf of publishers.
It replaces a prior work called \emph{Packaging on the Web}~\cite{packagingweb}.
Currently, its specification is split in three different modular layers, namely \emph{Signing}~\cite{wp:signing}, \emph{Bundling}~\cite{wp:bundling}, and \emph{Loading}~\cite{wp:loading}.
The \emph{Signed HTTP Exchanges} specification allows an origin to digitally sign one or more \emph{HTTP Exchanges} (an \emph{HTTP Exchange} is a pair of an \emph{HTTP Request} and corresponding \emph{HTTP Response}) so that they can be distributed on behalf of the origin by intermediaries while maintaining the authenticity of the content.
The \emph{Bundled HTTP Exchanges} specification describes how a group of one or more signed or unsigned \emph{HTTP Exchanges} from one or more origins can be bundled together for distribution.
The \emph{Loading Signed Exchanges} specification describes a set of algorithms to check whether a signature on an exchange is valid.
All three layers of \emph{Web Packaging} have the potential to play a significant role in web archiving.
They can help crawl resources more effectively, replay more accurately, and facilitate fixity and non-repudiation on archived resources.
However, realizing that potential requires some changes to the specification as it stands.
These changes include support for the \emph{Memento} protocol and long-lived trust of signed exchanges.
We urge the \emph{Web Packaging} community to consider how it can help to archive the web.\looseness=-1

\section{Web Packaging in Web Archiving}

Despite a wealth of activities, internationally, related to web archiving that started more than two decades ago, mainstream web systems and protocols have put insufficient emphasis on the need to be able to preserve web resources and access those preserved resources in the future.
The focus of technical advancements is on speed, efficiency, user experience, and security, but lacks in consideration of archivability and access to archived resources as a significant aspect.
This is disconcerting from a societal perspective, because without an archivable and archived web, revisiting the history of our era will be all but impossible~\cite{webhistory}.
When the \emph{Web Packaging} specification was announced, the web archiving community took notice~\cite{wpwarc:tweet} with the hope that it might help to mainstream web archiving.
Some others from web publishing and personal archiving backgrounds were also interested in utilizing \emph{Web Packaging} as a means to preserve content in an immutable manner with built-in long-term trust~\cite{immutable:gh}.
As a result an archival use case~\cite{archuse:gh} was added, but with the requirement of the content being unsigned~\cite{wpuse} to avoid expired signatures.
The remainder of this section describes the benefits \emph{Web Packaging} can bring to web archiving by facilitating effective crawling, accurate replay, and non-repudiation.\looseness=-1

\subsection{Effective Crawling With Bundled HTTP Exchanges}

With the proliferation of \emph{JavaScript} on the web it has become increasingly difficult to crawl web resources of a domain effectively and completely using traditional crawlers like \emph{Heritrix}~\cite{heritrix}.
Resources that do not appear in the plain \emph{HTML} or \emph{CSS} and are fetched only after client-side rendering, and possibly after a user interaction, are often not preserved~\cite{defrep}.
Large-scale crawlers maintain a frontier queue using data structures like priority queue and a set of recently seen \emph{URIs}.
This means some page requisites may be captured long after their parent pages and by then their state might have changed.
While there exist headless browser-based crawlers (e.g., \emph{Brozzler}\footnote{\url{https://github.com/internetarchive/brozzler}} and \emph{Squidwarc}\footnote{\url{https://github.com/N0taN3rd/Squidwarc}}), they are an order of magnitude or two slower than static crawlers.
\textbf{\emph{Bundled HTTP Exchanges} can be helpful in this case by serving a complete set of temporally coherent resources and saving the crawler from parsing a great deal of markup, assuming the server knows about all the requisites and bundles them effectively.}

\subsection{Coherent Replay With Bundled HTTP Exchanges}

Web archives serve \emph{mementos} on behalf of a different origin while the pages were designed with the original domain in mind.
This poses many difficulties in replaying a \emph{composite memento} correctly such as live-leakage~\cite{zombies}, temporal violations~\cite{tempcoherence}, origin violations~\cite{originviolations}, cookie violations~\cite{cookieviolations}, and broken links; all of which may yield a rendition of a page that never existed on the live web (e.g., a weather page saying sunny, but showing a rainy satellite image)~\cite{oneinfive} and some may pose security risks~\cite{rewritehist}.
Archival replay systems often perform extensive \emph{URL} rewriting to ensure that the subsequent page requisite requests are routed to the archives and not the live site or an invalid location.
\emph{URLs} that are generated by \emph{JavaScript} are difficult to identify and rewrite, resulting in broken \emph{composite mementos}.
Proxy or browser extension-based solutions exist to mitigate this, but they do not work out of the box and require users to configure their browsers.
Some replay systems use client-side rewriting (e.g., \emph{Wombat}~\cite{wombat}) or \emph{Service Worker}-based rerouting (e.g., \emph{Reconstructive}~\cite{reconstructive,ipwb}) to ensure that requests maintain the desired origin boundary.
The \emph{UK Web Archive} limits its replay to certain whitelisted sites that adhere to and advertise certain usage policies, and otherwise returns an \emph{HTTP 451} status code~\cite{http451:rfc}. It whitelists some domains like \texttt{twitter.com}, but fails to recognize its \emph{CDNs}, resulting in broken pages.

We envision a future in which web archives would have preserved signed or unsigned \emph{Bundled HTTP Exchanges} related to a requested \emph{composite memento} or could effectively identify all the resources needed (from one or more origins) for it to bundle them all in a single unsigned package with appropriate origin boundaries.
This means the user-agent would not need to resolve for any resources on the live web to render the \emph{composite memento}, thus avoiding many of the issues listed above.
This also means that archival replay systems would not need aggressive \emph{URL} rewriting and could serve originally preserved bytes on behalf of respective origins (except a few places where some rewriting might be inevitable).
We believe that effective use of \emph{Bundled HTTP Exchanges} can eventually solve many archival replay problems, resulting in temporally coherent and accurate \emph{composite mementos}.

It is worth noting that \emph{mementos} served from a web archive are ``a representation of a resource at a \emph{URI} \textbf{as observed at a given time in the past}'' instead of ``a representation of a resource at a \emph{URI}'', hence there might be many timestamped versions of the same resource in \emph{Bundles} and \emph{HTTP cache}.
Currently, \emph{Loading Signed Exchanges} favor the most recent version from a \emph{stashed exchange} or \emph{HTTP cache}, but in an archival context every version is equally as important.
\emph{Memento} compliant web archives resolve to a specific version of a resource when the \emph{TimeGate} associated with that resources receives a request with an \texttt{Accept-Datetime} header. 
The returned \emph{memento} has a \emph{Memento-Datetime} header to express the time when it was archived, as well as links pertaining to datetime negotiation in the \texttt{Link} header.
However, these additional headers and content negotiations are provisioned by an archival replay server and are not part of the original request and response (unless a web server is itself \emph{Memento} compliant).
In case of \emph{Signed Exchanges}, altering messages on the server side is not possible, hence any time-based content negotiation needs to be done on the client-side after the signature validation.
Alternatively, a \emph{Memento-Datetime} header can be returned with the \emph{Bundle} that can be used to namespace a cache and a special header to indicate resource resolution policy that tells the user-agent to not resolve a request if it is not present in the namespaced cache.
Such namespaced caches have an added advantage of creating a security boundary to limit some downgrade attacks if their access is tied to the origin of the bundle distributor.
\textbf{This means, to leverage \emph{Web Packaging} in web archiving to its full potential, \emph{Loading Signed Exchanges} and \emph{Fetch} algorithms need to be extended to support time-based content negotiation (i.e., built-in \emph{TimeGate} support within the \emph{Bundle}) for versioned resources to ensure resolution of the correct and temporally coherent version of resources.}

\subsection{Fixity and Non-repudiation With Signed Exchanges}

In order to use web archives in a legal environment it is essential to be able to prove that a \emph{memento} in question was not forged or altered (i.e., maintain \emph{fixity}) beyond what is necessary for proper replay and the content was indeed produced by the said origin (i.e., establish \emph{non-repudiation}).
Due to the lack of technical means of proving \emph{fixity} and \emph{non-repudiation} of \emph{mementos}, currently archive personnel has to certify \emph{fixity} when necessary (e.g., the case of Joy-Ann Reid claiming that copies of her blog in the Internet Archive has been hacked~\cite{joyreid:ia,joyreid:wsdl}).

Examining fixity of \emph{mementos} is difficult and often impossible in \emph{JavaScript}-rich \emph{composite mementos} due to inconsistencies in successive replays.
There are archival fixity proposals using web archives themselves~\cite{fixityblock} or a \emph{Blockchain}~\cite{archangel}, but they require ahead of time content digest advertisement and additional overhead of resources.
Moreover, these approaches can only track the fixity of a resource as advertised by a web archive, which can be different from what the origin of the resources has originally returned (i.e., lack of \emph{non-repudiation}).
In the event of an \emph{HTTPS} communication it sounds plausible that if original encrypted bits along with the complete \emph{TLS} handshake log were preserved, an archive should be able to establish \emph{non-repudiation}, but it can not, because \emph{HTTPS} traffic is encrypted using a shared key not an asymmetric one.
Consequently, while the archive itself can rest assured the response indeed came from the said origin, it has the ability to fake the \emph{TLS} handshake log, hence cannot prove the origin to anyone else.

This is where \emph{Signed Exchanges} can have an impact, but unfortunately, the trust of such signatures is short-lived, which is not suitable for archival time scale.
We see great technical potential in \emph{Web Packaging} of being helpful for web archiving if we can build a long-lasting history-aware temporal signature validation model.
By this we mean, rather than a digital signature being either ``\emph{valid}'' or ``\emph{invalid}'', introduce another state ``\emph{temporally valid}'', that indicates that the signature would have been valid at a given time in the past.
The \emph{Memento} framework already provides a standard means to express that a resource representation is historical, not live, using the \texttt{Memento-Datetime} header.
Using this, a user-agent would know that it needs to validate the signature in a temporal context and acknowledge the state visually (e.g., web browsers showing the state of a certificate in the address bar) or by some other means as suitable.
Fortunately, due to the rapid adoption of the tamper-proof and publicly auditable \emph{Certificate Transparency}~\cite{certtrans} by many certificate authorities, it seems possible to build a temporally-aware digital signature trust model.
It is worth noting that once a \emph{private key} is compromised, corresponding historical signatures will become ``\emph{invalid}'' too, because one can create back-dated fake records and sign them with the stolen key.

We feel that with the current short-lived validity of \emph{Signed Exchanges}, the \emph{Web Packaging} favors aggregators that are interested in the recent and live web (e.g., search engines, social media, and \emph{CDNs}), but hurts many culturally and historically important systems that require a trust system (both online and offline) that lasts long after the origins of the resources are gone (e.g., web archives, digital libraries, book readers, data sharing systems, and publications).
We do not think that this is intentional, rather a consequence of how our existing digital signature system on the web works.
\textbf{Hence, we believe that a temporal certificate validation extension would make \emph{Web Packaging} more inclusive and welcoming for entities at the lower end of the power graph.}\looseness=-1

\section{Conclusions}

We believe that the web archiving community is generally receptive of \emph{Web Packaging}, but would welcome changes that further increase its potential for web archiving and web archive access.
Web archiving is becoming increasingly challenging due to the rapid evolution of web technologies, but \emph{Web Packaging} can ameliorate some of those challenges with changes along the lines we described.
The \emph{Web Packaging} community has a unique opportunity to devise a technology that supports web archiving and provides a much needed capability to verify the integrity of archived web resources.
We hope it will decide to embrace this opportunity and we express our willingness to collaborate to make this happen.

\section{Acknowledgements}

This work is supported in part by the Andrew W. Mellon Foundation (AMF) grant 11600663.

\bibliographystyle{ACM-Reference-Format}
\bibliography{ref} 

\end{document}